\documentclass[sigconf,authorversion]{acmart}
\pdfoutput=1

\usepackage[utf8]{inputenc}

\usepackage{listings}
\usepackage{tikz}
\usepackage[nowarn]{glossaries}
\glsdisablehyper
\usepackage{pifont}
\usepackage{array}
\usepackage{multirow}
\usepackage{booktabs}
\usepackage{microtype}
\usepackage{xspace}
\PassOptionsToPackage{hyphens}{url}\usepackage{hyperref}

\newcommand{\Cpp}{C\texttt{++}\xspace}
\newcommand{\etal}{et al.\xspace}

\newcommand{\longjmp}{\texttt{longjmp()}\xspace}
\newcommand{\safelongjmp}{\texttt{safelongjmp()}\xspace}
\newcommand{\safesetjmp}{\texttt{safesetjmp()}\xspace}

\newcommand{\setjmp}{\texttt{setjmp()}\xspace}
\newcommand{\smmap}{\texttt{mmap()}\xspace}
\newcommand{\swrite}{\texttt{write()}\xspace}
\newcommand{\libc}{\texttt{libc}\xspace}
\newcommand{\libld}{\texttt{ld}\xspace}

\newcommand{\cmark}{\ding{51}}
\newcommand{\xmark}{\ding{55}}
\newcommand{\onemark}{\ding{182}}
\newcommand{\twomark}{\ding{183}}
\newcommand{\threemark}{\ding{184}}
\newcommand{\fourmark}{\ding{185}}

\setcopyright{acmlicensed}

\acmDOI{10.1145/3196494.3196531}

\acmISBN{978-1-4503-5576-6/18/06}

\acmConference[ASIA CCS '18]{2018 ACM Asia Conference on Computer and Communications Security}{June 4--8, 2018}{Incheon, Republic of Korea}
\acmBooktitle{ASIA CCS '18: 2018 ACM Asia Conference on Computer and Communications Security, June 4--8, 2018, Incheon, Republic of Korea}
\acmYear{2018}
\copyrightyear{2018}

\acmPrice{15.00}

\newacronym{ABI}{ABI}{Application Binary Interface}
\newacronym{ASLR}{ASLR}{Address Space Layout Randomisation}
\newacronym{CET}{CET}{Control-Flow Enforcement Technology}
\newacronym{CFG}{CFG}{Control-Flow Graph}
\newacronym{CFI}{CFI}{Control-Flow Integrity}
\newacronym{CPI}{CPI}{Code Pointer Integrity}
\newacronym{CPU}{CPU}{Central Processing Unit}
\newacronym{DEP}{DEP}{Data Execution Prevention}
\newacronym{DTV}{DTV}{Dynamic Thread Vector}
\newacronym{EAP}{EAP}{Ephemeral Allocation Primitive}
\newacronym{EH}{EH}{Exception Handling}
\newacronym{ELF}{ELF}{Executable and Linkable Format}
\newacronym{FP}{FP}{Frame Pointer}
\newacronym{IoT}{IoT}{Internet of Things}
\newacronym{ISA}{ISA}{Instruction Set Architecture}
\newacronym{JIT}{JIT}{Just-In-Time}
\newacronym{LAN}{LAN}{Local Area Network}
\newacronym{LR}{LR}{Link Register}
\newacronym{PAC}{PAC}{Pointer Authentication Codes}
\newacronym{PAP}{PAP}{Persistent Allocation Primitive}
\newacronym{PC}{PC}{Program Counter}
\newacronym{PCS}{PCS}{Procedure Call Standard}
\newacronym{PIC}{PIC}{Position-Independent Code}
\newacronym{PIE}{PIE}{Position-Independent Executable}
\newacronym{ROP}{ROP}{Return-Oriented Programming}
\newacronym{SP}{SP}{Stack Pointer}
\newacronym{TCB}{TCB}{Thread Control Block}
\newacronym{TID}{TID}{Thread ID}
\newacronym{TLS}{TLS}{Thread Local Storage}
\newacronym{W^X}{W$\oplus$X}{Write XOR Execute}

\makeglossaries

\begin{document}

\title[A Leak-Resilient Dual Stack Scheme for Backward-Edge CFI]{A Leak-Resilient Dual Stack Scheme for Backward-Edge Control-Flow Integrity}

\author{Philipp Zieris}
\affiliation{%
  \institution{Fraunhofer AISEC}
}
\email{philipp.zieris@aisec.fraunhofer.de}

\author{Julian Horsch}
\affiliation{%
  \institution{Fraunhofer AISEC}
}
\email{julian.horsch@aisec.fraunhofer.de}

\begin{abstract}
Manipulations of return addresses on the stack are the basis for a variety of attacks on programs written in memory unsafe languages.
Dual stack schemes for protecting return addresses promise an efficient and effective defense against such attacks.
By introducing a second, \emph{safe stack} to separate return addresses from potentially unsafe stack objects, they prevent attacks that, for example, maliciously modify a return address by overflowing a buffer.
However, the security of dual stacks is based on the concealment of the safe stack in memory.
Unfortunately, all current dual stack schemes are vulnerable to information disclosure attacks that are able to reveal the safe stack location, and therefore effectively break their promised security properties.
In this paper, we present a new, leak-resilient dual stack scheme capable of withstanding sophisticated information disclosure attacks.
We carefully study previous dual stack schemes and systematically develop a novel design for stack separation that eliminates flaws leading to the disclosure of safe stacks.
We show the feasibility and practicality of our approach by presenting a full integration into the LLVM compiler framework with support for the x86-64 and ARM64 architectures.
With an average of 2.7\% on x86-64 and 0.0\% on ARM64, the performance overhead of our implementation is negligible.

\end{abstract}

%
%



\maketitle

\vspace{1cm}

\section{Introduction}\label{sec:intro}

A vast majority of today's security-relevant vulnerabilities arise from the broad use of unsafe programming languages, such as C and \Cpp.
In favor of efficiency and flexibility, these languages omit the enforcement of strong type and memory safety.
The lack of such an enforcement frequently causes programming errors to result in vulnerable code pointers that can be corrupted at runtime in order to divert a program's control-flow and induce malicious program behavior~ \cite{SPWS13}.
A common form of runtime attacks on software takes control of its target by illicitly altering return addresses on the program's stack through a buffer overflow \cite{Sha07,CDD+10}.
The malicious program behavior is then carried out after the exploited function returns, diverting control to a location of the attacker's choosing.

This threat sparked the academic research of various defense techniques that commonly focus on securing the backward-edges---as taken by function returns---of a program's \gls{CFG}.
The first and most common approaches for backward-edge \gls{CFI}~\cite{ABEL05,DMW15} maintain copies of return addresses on a \emph{shadow stack} and use these copies to verify the integrity of function returns.
In order to reduce the performance overhead of shadow stacks, more
sophisticated solutions, namely SafeStack~\cite{KSP+14} and AG-Stack~\cite{LSL+15}, \emph{separate} potentially unsafe stack objects (e.g., local stack variables that store user-supplied data) from sensitive stack objects (e.g., return addresses).
They realize this as \emph{dual stack} schemes by introducing a second stack into the protected programs to hold the unsafe stack objects, while keeping the safe objects on the program's original stack.
These new stacks---commonly referred to as \emph{unsafe stack} and \emph{safe stack}---effectively thwart straightforward runtime attacks that overflow return addresses.
As they decouple vulnerable stack objects from the sensitive control-flow data, the attacker is unable to overwrite return addresses by exploiting the vulnerable objects.

However, recent runtime attacks~\cite{GGK+16,GKK+16,OABG16} undermine the security property of dual stacks.
Both SafeStack and AG-Stack rely on \emph{information hiding} and conceal the location of their safe stacks by means of \gls{ASLR}.
By using techniques to disclose the location of safe stacks, the attacker is able to overwrite return addresses using arbitrary vulnerable pointers within the program's memory (e.g., stack or heap), even if these return addresses are protected by backward-edge \gls{CFI} mechanisms.
Hence, utilizing information disclosure as part of a sophisticated runtime attack enables the attacker to circumvent the protective property of dual stack schemes altogether.
In detail, two major attack surfaces for information disclosures on dual stacks can be identified \cite{GGK+16}: (1)~Leaks of pointers to the safe stack, found in various, unsafe locations throughout the program's memory space, and (2)~the safe stack itself, prone to identification by searching the entire address space of the program.

In this paper, we present a new, leak-resilient dual stack design that
eliminates the aforementioned attack surfaces.
For this, we shift the paradigm of dual stacks from relocating potentially unsafe objects towards placing the sensitive objects on the new stack while keeping the unsafe objects on the program's original stack.
This basic design change enables our solution (1)~to completely avoid pointer leaks and (2)~to gain defensive properties against searching the safe stack within the program's address space.

We show the feasibility of our approach by presenting efficient implementations for the widely used x86-64 and ARM64 (AArch64) architectures.
To this end, we fully integrate our proposed dual stack design into the LLVM compiler framework and provide fully compatible Linux runtime environments on both architectures.
The changes applied to LLVM and the Linux runtime do not restrict any compiler and linker features, such as tail call optimization or \gls{PIC}, and are interoperable with language features, such as shortened return sequences (\setjmp/\longjmp), multi-threading, or \Cpp exceptions.
Further, the changes do not restrict the interoperability of protected programs with statically linked or dynamically loaded libraries, including unprotected legacy libraries.

In summary, we make the following contributions:
\begin{itemize}
  \item
    We propose a novel approach to stack separation that---by design---eliminates disclosure of the safe stack location by exploiting leaked pointers or searching the address space, using most recent techniques such as memory allocation oracles~\cite{OABG16}.
  \item
    We present a full integration of our proposed dual stack scheme into the LLVM compiler framework and the Linux runtime environment with complete support for the x86-64 and ARM64 architectures\footnote{Available at \url{https://github.com/llvm-return-stack}.}.
  \item
    We show that our approach is practically feasible with negligible performance overhead using standard CPU benchmarks and the Apache and Nginx web servers.
\end{itemize}
In the remainder of this paper, we introduce our threat model (\autoref{sec:threatmodel}), give a brief overview on information hiding (\autoref{sec:informationhiding}), study the weaknesses of previous dual stack designs (\autoref{sec:dualstacks}), describe the design of our leak-resilient dual stack scheme (\autoref{sec:design}), present our implementation (\autoref{sec:implementation}), evaluate the performance (\autoref{sec:evaluation}), discuss related work (\autoref{sec:rw}), and conclude (\autoref{sec:conclusion}).

\section{Threat Model}\label{sec:threatmodel}

Dual stack schemes are designed to prevent runtime attacks that achieve malicious code execution by exploiting backward-edges in a program's \gls{CFG}.
To evaluate our own dual stack approach and draw comparisons to other approaches, we define a threat model within which the ultimate goal of an attacker is to maliciously alter a return address on the target program's stack.
We base our threat model on prior work in the area \cite{ABEL05,TRC+14,LSL+15}.

First, we assume that the compiler toolchain is trusted and reliably inserts our defense mechanism into the target program.
Next, we assume that the target program is running on a system where the hardware and all privileged software components, such as hypervisors or operating systems, are trusted and operating correctly.
Hence, program-specific runtime data held by privileged components (e.g., register states) is out of the attacker's reach.
The system further employs standard defense mechanisms, i.e., \gls{DEP}, \gls{W^X}, and \gls{ASLR}.
Therefore, the target's program code and read-only data cannot be modified by the attacker and \gls{JIT} compiled and self-modifying code are out of scope.

We assume that the target program contains memory errors resulting in (remotely) exploitable, input-controlled vulnerabilities granting the attacker the following capabilities:
\begin{itemize}
	\item
    The attacker is able to \emph{read from any memory location} in the target's address space.
    This enables him to perform information disclosure attacks in order to leak the process' memory layout.
	\item
    The attacker is able to \emph{write data to any writable location} in the target's address space.
		This includes \emph{non-contiguous} and \emph{contiguous}
		writes. The latter describes a write to a location directly adjacent to a
		vulnerable location, for example, a buffer overflow.
\end{itemize}
These assumptions yield a realistic threat model with strong attacker capabilities.
The goal of our attacker is to alter a return address on the target program's stack granting malicious code execution, for example, in form of a chain of return-oriented instruction sequences (i.e., gadgets) or a function call with attacker-supplied parameters (e.g., \texttt{execve()} to spawn a shell).

\section{Information Hiding}\label{sec:informationhiding}

Dual stack schemes, like most modern defenses that protect the control-flow of programs, rely on \emph{information hiding} to strengthen their protective properties.
As such, these defenses separate sensitive pointers (e.g., return addresses, function pointers, and dispatch tables) and metadata thereof from everything else in memory by placing them in self-contained \emph{hidden areas}.
Access to these areas is then restricted to legitimate and non-exploitable code, concealing the areas from unauthorized access by attackers.
However, new attacks circumvent this concealment by means of \emph{information disclosure}, i.e., by searching for the hidden areas and leaking their locations.
To conduct such an information disclosure, the attacker has to overcome the entropy with which the hidden area is placed in the target process' address space.
Depending on the defense, this entropy is equal to the size of the process' address space or the entropy introduced by the deployed information hiding technique.

Our leak-resilient dual stack, as the previous dual stack schemes SafeStack
and AG-Stack, is developed for the Linux operating system.
On Linux, the common technique deployed for information hiding is \gls{ASLR}.
In the following, we discuss the virtual memory layout of a typical, 64-bit Linux (kernel version 4.14.4) process under \gls{ASLR}.
We highlight the differences between the \gls{ASLR} implementations on the x86-64 and ARM64 architectures and briefly discuss their implications on the security of dual stacks and other defenses.

\begin{figure}
  \centering
  \begin{tikzpicture}
  %
  %
  \usetikzlibrary{arrows.meta, decorations, decorations.pathreplacing, positioning, fit, positioning, shadows.blur}
  %
  %
  \tikzstyle{label}=[
    draw=none,
    inner sep=0pt,
    node distance=9pt, outer sep=0pt,
    font=\ttfamily\small
  ]
  \tikzstyle{area}=[ 
    draw=none,
    inner sep=0pt,
    minimum width=2.2cm, minimum height=.7cm,
    node distance=0pt, outer sep=0pt,
    font=\small, align=center, text width=2.2cm
  ]
  \tikzstyle{mapped}=[ 
    area, fill=black!15
  ]
  \tikzstyle{unmapped}=[ 
    area, fill=white!100
  ]
  \tikzstyle{invisible}=[
    draw=none, fill=none
  ]
  \tikzstyle{sbox}=[ 
    draw=black!100, thick,
    fill=black!40,
    inner sep=0pt,
    blur shadow
  ]
  \tikzstyle{line}=[
    draw=black!100, thin
  ]
  \tikzstyle{dots}=[
    line, dotted
  ]
  \tikzstyle{brace}=[
    line, decorate,
    decoration={brace, amplitude=3pt, mirror}
  ]
  \tikzstyle{growth}=[
    line, ->, >=Stealth
  ]
  \tikzstyle{every shadow}=[ 
      shadow opacity=60,
      shadow blur radius=.8ex,
      shadow xshift=0ex,
      shadow yshift=-.5ex,
      shadow blur steps=50,
      shadow blur extra rounding
  ]
  %
  %
  \pgfdeclarelayer{frame} 
  \pgfdeclarelayer{sb} 
  \pgfsetlayers{frame,sb,main}
  %
  %
  \node[unmapped]                      (stack_gap)  {};
  \node[mapped,   below=of stack_gap]  (stack_area) {Stack};
  \node[unmapped, below=of stack_area] (mmap_gap)   {};
  \node[mapped,   below=of mmap_gap]   (mmap_area)  {\baselineskip=8pt Memory mapping\\ space\par};
  \node[unmapped, below=of mmap_area]  (big_gap)    {};
  \node[mapped,   below=of big_gap]    (heap_area)  {Heap};
  \node[unmapped, below=of heap_area]  (heap_gap)   {};
  \node[mapped,   below=of heap_gap]   (elf_area)   {\baselineskip=8pt Loadable \glstext{ELF}\\ segments\par};
  \node[unmapped, below=of elf_area]   (elf_gap)    {};
  \draw[line] (stack_gap.south west)  -- (stack_gap.south east);
  \draw[line] (stack_area.south west) -- (stack_area.south east);
  \draw[line] (mmap_gap.south west)   -- (mmap_gap.south east);
  \draw[dots] (mmap_area.south west)  -- (mmap_area.south east);
  \draw[dots] (big_gap.south west)    -- (big_gap.south east);
  \draw[line] (heap_area.south west)  -- (heap_area.south east);
  \draw[line] (heap_gap.south west)   -- (heap_gap.south east);
  \draw[line] (elf_area.south west)   -- (elf_area.south east);
  \draw[line] (elf_gap.south west)    -- (elf_gap.south east);
  \draw[line]   ([xshift=3pt]stack_gap.north east) -- +(4pt,0pt);
  \draw[line]   ([xshift=3pt]stack_area.south east) -- +(4pt,0pt);
  \draw[line]   ([xshift=3pt]elf_gap.south east) -- +(4pt,0pt);
  \draw[line]   ([xshift=3pt]mmap_area.north east) -- +(4pt,0pt);
  \draw[growth] ([xshift=5pt]mmap_area.north east) -- +(0pt,-.8cm);
  \draw[line]   ([xshift=3pt]heap_area.south east) -- +(4pt,0pt);
  \draw[growth] ([xshift=5pt]heap_area.south east) -- +(0pt,.8cm);
  \node[label, right=of stack_gap.north east]  {$2^b$};
  \node[label, right=of stack_area.south east] {stack\_base};
  \node[label, right=of mmap_area.north east]  {mmap\_base};
  \node[label, right=of mmap_area.south east]  {};
  \node[label, right=of heap_area.north east]  {brk};
  \node[label, right=of heap_area.south east]  {brk\_start};
  \node[label, right=of elf_gap.south east]    {0x0};
  \draw[brace] ([xshift=-3pt]stack_gap.north west) -- ([xshift=-3pt]stack_gap.south west);
  \draw[brace] ([xshift=-3pt]mmap_gap.north west)  -- ([xshift=-3pt]mmap_gap.south west);
  \draw[brace] ([xshift=-3pt]heap_gap.north west)  -- ([xshift=-3pt]heap_gap.south west);
  \draw[brace] ([xshift=-3pt]elf_gap.north west)   -- ([xshift=-3pt]elf_gap.south west);
  \node[label, left=10pt of stack_gap] {stack\_offset};
  \node[label, left=10pt of mmap_gap]  {mmap\_offset};
  \node[label, left=10pt of heap_gap]  {brk\_offset};
  \node[label, left=10pt of elf_gap]   {load\_offset};
  \begin{pgfonlayer}{sb}
    \node[sbox, fit=(stack_gap)(elf_gap)] (stack_shadow) {};
  \end{pgfonlayer}
  \begin{pgfonlayer}{frame}
      \node[inner sep=.2cm, fit=(stack_gap)(elf_gap)] {};
  \end{pgfonlayer}
\end{tikzpicture}
  \caption{Memory layout of an \glstext{ASLR}-enabled process on 64-bit Linux.}
  \label{fig:memory_layout}
\end{figure}
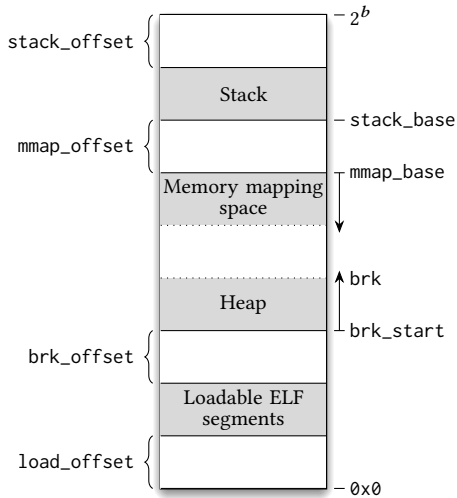

Processes are provided a virtual address space of $2^b$ bytes by the Linux kernel, organized in virtual pages of 4 KiB ($2^{12}$ bytes) and spanning over the address range $[0,2^b)$.
On the x86-64 architecture, the number of address bits $b$ is 47, yielding an
address space of 128 TiB ($2^{47}$ bytes; $2^{35}$ pages). On ARM64 $b$ is 48, yielding an address space of 256 TiB ($2^{48}$ bytes; $2^{36}$ pages).
The kernel populates the address space deterministically, starting from the lowest address, with the loadable \glstext{ELF} segments (\texttt{.text}, \texttt{.data}, \texttt{.bss}, etc.) followed by the heap, and, from the highest address, with the stack followed by the memory mapping space.
Accordingly, the stack and memory mapping space grow down towards lower addresses, while the heap grows in the opposite direction.
The layout of the virtual address space is depicted in \autoref{fig:memory_layout}.
The memory mapping space is organized by the kernel's \smmap infrastructure and initially populated with shared libraries that are required during the process' initialization.
These libraries are loaded in a deterministic manner and typically include \libld and \libc.
During execution, the memory mapping space also holds shared libraries loaded dynamically as needed, (large) heap allocations backed by \smmap, as well as file and anonymous mappings retrieved through \smmap.

During the initialization of the process, the kernel randomizes the memory layout by applying randomly chosen offsets to the base addresses of the loadable \glstext{ELF} segments, the heap, the memory mapping space, and the stack.
The offset to the loadable \glstext{ELF} segments is only randomized for \glspl{PIE} and otherwise fixed to a pre-defined value.
The different intervals, the offsets are randomly chosen from, are illustrated for both architectures in \autoref{tab:randomization_intervals}.
For each interval, the offset is page aligned, reducing the available bits for randomization (i.e., the entropy) by the page size.
The addressable bits and resulting entropy for each interval are illustrated in \autoref{tab:randomization_intervals} as well.

Typically, defenses that rely on information hiding place their hidden areas within the memory mapping space.
This space grows towards the heap and can occupy a vast majority of the process' available virtual memory.
These defenses are therefore able to allocate large hidden areas, offering an entropy of 28 bits on x86-64 but only 18 bits on ARM64.
Other defenses, such as SafeStack and AG-Stack, that declare the stack as a
hidden area, only achieve an entropy of 22 bits and 18 bits on x86-64 and ARM64.

\begin{table}
  \caption{Randomization intervals of \glstext{ASLR} and their resulting entropies on 64-bit Linux for the x86-86 and ARM64 architectures.}
  \label{tab:randomization_intervals}
  \begin{tabular}{cllcc}
    \toprule
    & \textbf{Name} & \textbf{Interval}\footnotemark & \textbf{Bits} & \textbf{Entropy} \\
    \midrule
    \parbox[t]{4pt}{\multirow{4}{*}{\rotatebox[origin=c]{90}{\textbf{x86-64}}}}
    & \texttt{stack\_offset} & $[\texttt{0},\texttt{0x400000000})$   & 34 & 22\\
    & \texttt{mmap\_offset}  & $[\texttt{0},\texttt{0x10000000000})$ & 40 & 28\\
    & \texttt{brk\_offset}   & $[\texttt{0},\texttt{0x2000000})$     & 25 & 13\\
    & \texttt{load\_offset}  & $[\texttt{0},\texttt{0x10000000000})$ & 40 & 28\\
    \midrule
    \parbox[t]{4pt}{\multirow{4}{*}{\rotatebox[origin=c]{90}{\textbf{ARM64}}}}
    & \texttt{stack\_offset} & $[\texttt{0},\texttt{0x40000000})$ & 30 & 18\\
    & \texttt{mmap\_offset}  & $[\texttt{0},\texttt{0x40000000})$ & 30 & 18\\
    & \texttt{brk\_offset}   & $[\texttt{0},\texttt{0x40000000})$ & 30 & 18\\
    & \texttt{load\_offset}  & $[\texttt{0},\texttt{0x40000000})$ & 30 & 18\\
    \bottomrule
  \end{tabular}
\end{table}

\footnotetext{Taken from \texttt{randomize\_stack\_top()}, \texttt{arch\_pick\_mmap\_layout()}, \texttt{arch\_ran\-dom\-ize\_brk()}, and \texttt{load\_elf\_binary()}, respectively.}

\section{On the Security of Dual Stacks}\label{sec:dualstacks}

In recent years, dual stack schemes emerged as the most sophisticated defense mechanism against attackers focusing on exploiting backward-edges in a program's \gls{CFG}.
Dual stack schemes divide a program's stack into two separated areas, the \emph{unsafe stack} holding potentially unsafe stack objects, such as local stack variables that store user-supplied data, and the \emph{safe stack} holding sensitive stack objects, such as return addresses that are crucial to the integrity of the program's control-flow.
To strengthen the resilience against information disclosure attacks, these schemes randomize the location of the safe stack within the address space of the protected program.
The two main dual stacks approaches are SafeStack~\cite{KSP+14} and AG-Stack~\cite{LSL+15}.
Before presenting our leak-resilient dual stack design, we introduce both mechanisms below and discuss recent advances in targeted information disclosure attacks against them.

\subsection{SafeStack}\label{sec:dualstacks_safestack}

SafeStack was initially presented by Kuznetsov \etal~\cite{KSP+14} as part of a general solution to enforce \gls{CPI}.
Yet, SafeStack works independently from the overall \gls{CPI} solution and can be used to guarantee backward-edge \gls{CFI}.
As a result, SafeStack has been integrated into the LLVM compiler framework as an architecture-independent\footnote{The extension supports the ARM, ARM64, MIPS32, MIPS64, x86, and x86-64 architectures.} extension~\cite{LLVM-SafeStack}.

SafeStack performs compile-time analysis in order to distribute all stack objects onto a safe and an unsafe stack.
This analysis considers register spills and return addresses to be always safe.
The safety of function parameters and local variables is determined by analyzing their type and scope.
Stack objects are deemed safe if they are accessed exclusively within their corresponding function and only with a fixed offset through the local stack pointer or frame pointer.
All remaining stack objects, i.e., those passed to child functions, are marked unsafe and stored on the unsafe stack.

In regard to the implementation of SafeStack, both the original work and the LLVM extension save the reference to the safe stack in the program's stack pointer register (e.g., \texttt{RSP} on x86-64) and the reference to the unsafe stack in the program's \gls{TLS}, as multi-threading is supported.
Both implementations use memory mappings to store their unsafe stacks and dereference the pointer from the \gls{TLS} upon access to unsafe stack objects.
The two implementations differ in terms of the placement of safe stacks.
Kuznetsov \etal store safe stacks within a (much) larger memory mapping---the safe region---used by the overall \gls{CPI} solution to store sensitive data.
In contrast, due to the lack of this region, the LLVM extension opts to store safe stacks as a replacement for the program's original stacks.
As a result, the main thread's original stack becomes the safe stack and each child thread's safe stack is allocated as a dedicated memory mapping.

\subsection{AG-Stack}\label{sec:dualstacks_agstack}

AG-Stack was presented by Lu \etal~\cite{LSL+15} as part of a general solution---called \glstext{ASLR}-Guard---to prevent information disclosure attacks that leak the location of any executable program code or any pointer to executable program code, ultimately rendering code-reuse attacks infeasible.
Lu \etal achieve this goal by separating and randomizing code and data sections of a program, encrypting pointers to program code when they are treated as data, and deploying a dual stack scheme---namely AG-Stack---to secure return addresses.
Hence, within \glstext{ASLR}-Guard, AG-Stack provides backward-edge \gls{CFI}.

AG-Stack distributes stack objects onto a safe and an unsafe stack according to a straightforward approach: The only objects considered to be safe are return addresses.
All other objects on a stack, namely register spills, function parameters, and local variables are deemed unsafe and stored on the unsafe stack.
The prototype implementation by Lu \etal was conducted on the x86-64 architecture and also supports multi-threading.
AG-Stack keeps its references to the safe and unsafe stack in two dedicated registers, namely \texttt{RSP} and \texttt{R15}.
Further, like SafeStack, AG-Stack stores the unsafe stacks as memory mappings and uses the safe stacks as a replacement for the program's original stacks.

\subsection{Information Disclosure Attacks}

Both SafeStack and AG-Stack conceal their safe stacks within the address space of the protected program.
For this, they rely on the principles of information hiding, and are therefore prone to information disclosure attacks.
These attacks exploit memory corruptions, i.e., use vulnerable data variables or data pointers \cite{SPWS13}, to locate a hidden area, such as the safe stack, within a much larger memory space.
In recent years, attackers developed different approaches to locate hidden areas by means of leveraging leaked pointers in linked libraries, probing the program's address space in a brute-force manner, and utilizing memory allocation oracles.
We discuss these attacks briefly in the following, before describing our dual stack design, which withstands \emph{all} of them.

As described in \autoref{sec:dualstacks_safestack}, the original SafeStack scheme places its safe stacks within the \gls{CPI} safe region.
Thus, information disclosure attacks against these stacks require leaking the safe region, which imposes different obstacles on the attacker.
Hence, we base our discussion regarding SafeStack on its implementation in LLVM.
Further, as described in \autoref{sec:dualstacks_agstack}, AG-Stack is part of the overall defense solution \glstext{ASLR}-Guard.
Since we do not have access to an implementation of \glstext{ASLR}-Guard, we base our discussion of AG-Stack solely on the information given in the paper by Lu \etal\cite{LSL+15}.

\subsubsection{Pointer Leaks}\label{sec:dualstacks_attacks_leaks}

A major aspect in keeping hidden areas concealed is the protection of references to these areas.
Typically, \gls{CFI} solutions realize this by accessing the hidden areas exclusively through dedicated registers.
However, even assuming protected programs themselves never write these dedicated registers to unprotected memory, it is not guaranteed that libraries linked into the programs do not leak them anyhow.
Therefore, a valid approach for an attacker to disclose hidden areas is to leverage such pointer leaks created at known memory locations by linked libraries.

Considering dual stack schemes, leaks of the stack pointer register (e.g., \texttt{RSP} on x86-64) reveal the location of safe stacks, in case the schemes utilize that register to hold the safe stack pointer (like SafeStack and AG-Stack).
The stack pointer register is typically written to memory by libraries that intervene with the control-flow of programs.
On GNU/Linux runtime environments, most prominently, the GNU C standard and GCC runtime libraries perform control-flow changes due to features such as shortened return sequences, stack unwinding, and multi-threading.

\paragraph{Shortened Return Sequences}

Through the \setjmp/\longjmp interface, the GNU C standard library allows functions to return to a different function than their immediate caller.
This creates a shortened return sequence, as functions in-between the call site of \setjmp and the invocation of \longjmp are not returned to, but skipped entirely.
For this, a copy of the entire register state is created upon entering \setjmp and reinstated into the registers by invoking \longjmp.
The register state is stored at a programmer-chosen memory location, potentially leaking the stack pointer register to unsafe memory (e.g., when the register state is stored on the heap).

\paragraph{Stack Unwinding}

The GCC runtime library implements stack unwinding to facilitate \Cpp exception handling, debugging functionalities, and multi-threading.
During unwinding, the unwinder traverses backwards through the stack cleaning up stack frames along the way.
To keep track of the stack frames, the unwinder stores the stack pointer of the current frame in an unwinding context.
This unwinding context is allocated on the heap creating a leak of the stack pointer register to unsafe memory.

\paragraph{Multi-threading}

Support for multi-threading is provided by the GNU C standard library in form of POSIX threads.
On creation of new threads, the library allocates an extra stack in the memory mapping space and stores its base address and size within a corresponding \gls{TCB}.
References to this \gls{TCB} are then passed---via the thread ID---to programmer-controlled variables or stored at known offsets in the library's data segment (e.g., \texttt{stack\_used}).
Thus, the location of the stack is leaked at multiple locations throughout the unsafe memory.

\paragraph{Impact on SafeStack and AG-Stack}

Both schemes are affected by pointer leaks in linked libraries.
Because the underlying causes of these leaks differ, two types can be distinguished:
\begin{description}
  \item[Neglected Spills.]
    Libraries that directly intervene with the control-flow of programs need to store register states to memory, spilling the register dedicated to hold the safe stack pointer (e.g., shortened return sequences and stack unwinding).
  \item[Neglected Metadata.]
    Libraries that maintain stack metadata and keep references to that metadata throughout their memory leak the safe stacks of SafeStack and AG-Stack by default, because both schemes replace the process' original stacks with safe stacks (e.g., multi-threading).
\end{description}
SafeStack and AG-Stack do not deploy any form of countermeasures to prevent pointer leaks through spills and metadata.

\subsubsection{Brute-force Probing}\label{sec:dualstacks_attacks_probing}

Another approach to disclose hidden areas is scanning the entire address space of the targeted program in a brute-force manner, and, at the same time, probe each encountered page whether it belongs to the hidden area.
When conducting such a scan, the attacker has to overcome the entropy of the searchable address space, which, depending on the implementation, is equal to the size of the addressable memory space or the entropy introduced by the used information hiding technique.

However, attackers have managed to drastically reduce this entropy by leveraging information about the hidden areas, such as its size, internal structure (i.e., sparsely populated or characteristic byte sequences), and placement within the larger address space (i.e., vicinity to known mappings, such as libraries).
Like this, recent attacks successfully disclosed the safe region of \gls{CPI}~\cite{EFG+15,GKK+16} and the safe stacks of the LLVM extension~\cite{GGK+16}.
These attacks conducted brute-force searches by utilizing fault analysis and
timing side-channels, both of which originally used to search for gadgets
(e.g., within shared libraries) in \glstext{ASLR}-protected programs to
facilitate code-reuse attacks~\cite{BBM+14,SOS14}.

Göktaş \etal \cite{GGK+16} studied the effectiveness of fault-tolerant memory oracles to locate small-sized hidden areas, and, in the process, successfully attacked and disclosed the safe stacks of LLVM.
In LLVM, safe stacks are created on a per-thread basis, with each stack being at most 8 MiB ($2^{23}$ bytes; $2^{11}$ pages) in size and allocated in the memory mapping space.
With an entropy of 28 bits\footnote{These values are for the x86-64 architecture. On ARM64, the entropies are even lower.} for memory-mapped addresses, this results in $ 2^{17}$ ($2^{28}/2^{11}$) possible page-aligned start addresses for each safe stack.

In order to reduce this entropy, Göktaş \etal developed two techniques---namely \emph{thread spraying} and \emph{stack spraying}---that artificially prepare the safe stacks for attacks.
With thread spraying, the authors exploit (mis-)configurations of target programs that allow spawning a multitude of worker threads (e.g., with JavaScript control in browsers to start web workers at will).
For each new thread, a safe stack is allocated in the memory mapping space using \smmap.
As subsequent calls to \smmap allocate memory at consecutive addresses, fast successively spawned threads create a contiguous memory region containing safe stacks only.
Using this approach, Göktaş \etal spawn 200 worker threads and increase the combined stack size to 400 MiB ($2^{16}$ pages), in turn reducing the search area to $2^{28}/2^{16}=2^{12}$ pages.
To find safe stacks within this area, Göktaş \etal further conduct stack spraying to populate the stacks with unique byte sequences that are easily searchable.
For this, they utilize web workers within their controlled JavaScript environment to call a function recursively with their unique byte sequence as argument.
Thus, the memory location of a safe stack can be obtained by probing at worst $2^{12}$ pages for the byte sequence.

\paragraph{Impact on SafeStack and AG-Stack}

Both schemes are susceptible to brute-force probing due to several reasons.
First, as both schemes replace the process' original stacks with safe stacks, the main thread's safe stack receives an entropy of only 22 bits, while the safe stacks of child threads receive an entropy of 28 bits.
As safe stacks are allocated with a default (maximum) stack size of 8 MiB, the effective entropy for a linear search is reduced to 11 bits ($2^{22}/2^{11}$) for the main thread and 17 bits ($2^{28}/2^{11}$) for all child threads.

Second, due to the allocation through \smmap, the safe stacks of child threads are likely to be placed in memory locations adjacent to each other or to well-known libraries.
This can further reduce the effective entropy, as the safe stacks form a contiguous memory region stretching over a multiple of $2^{11}$ pages and the adjacency to libraries enables an attacker to search for these libraries (which ideally occupy more than $2^{11}$ pages) instead of the safe stacks.

Finally, both schemes are inherently prone to thread spraying and SafeStack is additionally exposed to stack spraying.
For both, thread spraying is amplified by the fact that memory mappings retrieved through \smmap observe a spatial adjacency.
Stack spraying, on the other hand, is thwarted by AG-Stack, as return addresses are strictly separated from other stack objects by placing them exclusively on the safe stack.
In doing so, an attacker is not able to exploit other stack objects to artificially populate the safe stacks.

\subsubsection{Memory Allocation Oracles}\label{sec:dualstacks_attacks_mac}

Instead of probing memory in a brute-force manner, hidden areas can also be disclosed with constant costs using memory allocation oracles.
Memory allocation oracles reveal the size of holes (unmapped memory) between allocations inside the target's address space.
Therefore, by identifying all holes in an address space, hidden areas can be trivially inferred by an attacker.
Oikonomopoulos \etal \cite{OABG16} presented attacks based on memory
allocation oracles and broke all current \gls{CFI} and \gls{CPI} solutions that rely on information hiding, including the \gls{CPI} safe region, the safe stacks of LLVM, and \glstext{ASLR}-Guard with AG-Stack.

In their paper, Oikonomopoulos \etal describe two types of memory allocation oracles: \glspl{EAP} that create short-lived allocations and \glspl{PAP} that create allocations with a lifetime lasting for the duration of the attack.
To craft these primitives, the authors rely on target programs that allocate memory objects (e.g., using \smmap) as part of their input-handling logic, and additionally on a memory corruption that grants an attacker the ability to control the size of these memory allocations.
Using the example of web servers, when an attacker is able to allocate memory as part of HTTP client connections, an \gls{EAP} can be crafted using a non-persistent connection, while a \gls{PAP} can be crafted using a persistent connection.

An attacker with access to an \gls{EAP} and its size parameter can perform a binary search to find the larger one of the two holes surrounding the hidden area.
In each binary search iteration, the attacker performs a single EAP invocation and observes the positive (successful allocation; empty space) or negative (unsuccessful allocation; space contains parts of the hidden area) feedback.
Based on this feedback, the attacker adapts the size for the next iteration.
At the end, the search naturally returns the size of the larger hole.
To generalize the attack, i.e., searching for multiple holes within an address space, the attacker combines \gls{EAP} with \gls{PAP} in order to permanently allocate holes already discovered.

\paragraph{Impact on SafeStack and AG-Stack}

Information disclosure attacks based on memory allocation oracles reveal the entire memory layout of processes, enabling an attacker to distinguish between mapped and unmapped memory.
As both dual stack schemes allocate safe stacks as self-contained memory objects, their locations can trivially be inferred when the memory layout is known.

\subsection{Summary}\label{sec:dualstacks_summary}

Both SafeStack and AG-Stack are vulnerable to all three types of information disclosure, as summarized in \autoref{tab:dualstack_resilience}.
For brute-force probing, the overall vulnerability is comprised of the effective entropy (distinguishable for main/child threads and dependent on \glstext{ASLR} entropy and maximum stack size), spatial adjacency (received through \smmap) and general susceptibility to thread and stack spraying.
\begin{table}
  \caption{Resilience of dual stack schemes against information disclosure attacks.}
  \label{tab:dualstack_resilience}
  \begin{tabular}{l>{\centering\arraybackslash}m{1.0cm}>{\centering\arraybackslash}m{1.0cm}>{\centering\arraybackslash}m{2.1cm}}
    \toprule
    & \textbf{Safe\-Stack} & \textbf{AG-Stack} & \textbf{Leak-Resilient Dual Stack} \\
    \midrule
    \textbf{Pointer Leaks} & \xmark & \xmark & \cmark \\
    \midrule
    \textbf{Brute-Force Probing} & & & \\
    (ASLR) Entropy & 22/28 & 22/28 & 32 \\
    Max. Size (in pages) & $2^{11}$ & $2^{11}$ & $2^3$ \\
    Effective Entropy & 11/17 & 11/17 & 29 \\
    Spatial Adjacency & \xmark & \xmark & \cmark \\
    Thread Spraying & \xmark & \xmark & \xmark \\
    Stack Spraying & \xmark & \cmark & \cmark \\
    \midrule
    \textbf{Allocation Oracles} & \xmark & \xmark & \cmark \\
    \bottomrule
  \end{tabular}
\end{table}

The weaknesses of SafeStack and AG-Stack can be attributed to four specific design flaws:
\begin{description}
  \item[\onemark\ Structural Flaw.]
    Safe stacks that contain additional objects besides return address are susceptible to stack spraying (this only applies to SafeStack, not AG-Stack).
  \item[\twomark\ Oversizing.]
    The size of safe stacks directly weakens the resilience against brute-force probing by reducing the effective entropy.
    While dual stack schemes divide stack objects onto two stacks, both SafeStack and AG-Stack allocate safe stacks with the system-default size of (at most) 8 MiB ($2^{11}$ pages).
  \item[\threemark\ Architectural Conformity.]
    When replacing the process' original stacks with safe stacks and using the architecture's regular stack pointer register (i.e., \texttt{RSP} for SafeStack and AG-Stack) to reference these stacks, they are susceptible to pointer leaks through neglected metadata.
  \item[\fourmark\ Inherited Inflexibility.]
    Due to the architectural conformity, safe stacks also have to adhere to the system's specifications.
    For this reason, safe stacks strictly receive the entropy of \texttt{stack\_offset} and \texttt{mmap\_offset} (see \autoref{fig:memory_layout}), and further inherit the weaknesses imposed by \smmap, i.e., a spatial adjacency in memory, as well as a susceptibility to thread spraying and memory allocation oralces.
\end{description}
Furthermore, both SafeStack and AG-Stack suffer from pointer leaks through neglected spills, which cannot be avoided by a specific design choice but must be handled manually as described in \autoref{sec:implementation_linux}.
An illustration of the design flaws is given in \autoref{tab:design_flaws}.
Based on these flaws, we will present the design differences of our leak-resilient dual stack in the following section.

\begin{table}
  \caption{Design flaws of SafeStack and AG-Stack. The numbers correspond to \onemark\ the structural flaw, \twomark\ oversizing, \threemark\ the architectural conformity, and \fourmark\ the inherited inflexibility.}
  \label{tab:design_flaws}
  \begin{tabular}{lcccc}
    \toprule
    & \multicolumn{4}{c}{\textbf{Design Flaws}}\\
    \midrule
    \textbf{Pointer Leaks} & & & \large\threemark &\\
    \midrule
    \textbf{Brute-Force Probing} & & & &\\
    Effective Entropy & & \large\twomark & & \large\fourmark\\
    Spatial Adjacency & & & & \large\fourmark\\
    Thread Spraying & & & & \large\fourmark\\
    Stack Spraying & \large\onemark & & &\\
    \midrule
    \textbf{Allocation Oracles} & & & & \large\fourmark\\
    \bottomrule
  \end{tabular}
\end{table}

\section{Leak-Resilient Dual Stack Design}\label{sec:design}

According to our threat model from \autoref{sec:threatmodel}, dual stack schemes have to protect return addresses against contiguous and arbitrary writes to enforce backward-edge \gls{CFI} effectively.
The prevention of contiguous writes is achieved through the basic design principle of dual stacks: Separating potentially unsafe stack objects, such as local stack variables storing user-supplied data, from sensitive stack objects, i.e., the return addresses.
Hence, we maintain this basic design and separate stack objects onto safe and unsafe stacks.

To prevent arbitrary writes on safe stacks, their locations must be kept secret from an adversary.
Therefore, the safe stacks have to withstand powerful information disclosure attacks that, according to our treat model, are capable of reading arbitrary memory locations within the target program's address space.
In order to harden dual stack schemes, we improve on the four design flaws identified in \autoref{sec:dualstacks} and summarized in \autoref{tab:design_flaws}.
In the following, we present the improvements to these flaws and build our leak-resilient dual stack scheme.
The improvements in terms of leak-resilience achieved with our dual stack scheme are summarized in \autoref{tab:dualstack_resilience}.

\subsection{Return Stacks}\label{sec:design_stacks}

In order to prevent structural flaws (\onemark), an important design choice for dual stack schemes is the differentiation between safe and unsafe stack objects.
In this regard, and as discussed in \autoref{sec:dualstacks_attacks_probing}, the naïve approach of declaring all stack objects, except return addresses, unsafe is more effective against attacks that conduct stack spraying.
The lack of controllable objects prevents an attacker from artificially populating the safe stack to his own choosing, taking away his ability to identify safe stacks easily during brute-force probing.
Thus, we adopt this design choice and only store return addresses on our safe stacks, which we name \emph{return stacks} accordingly.

\subsection{Return Stack Size}\label{sec:design_size}

Capitalizing on the fact that we only store return addresses on our return stacks, we are able to counteract the oversizing flaw~(\twomark).
The size required by the return stack directly depends on the number of nested function calls in the protected program.
We measured the maximum function call depth for SPEC CPU2017 by instrumenting function prologues and epilogues.
The results, summarized in \autoref{tab:design_fcd}, show an average call depth of 465 and a maximum of 2581 for the \texttt{gcc} benchmark.
As one virtual memory page is able to hold 512 64-bit addresses, we conclude that a return stack size of 8 pages is sufficient for practical use.

\begin{table}
  \caption{Maximum function call depth of the SPEC CPU2017 benchmark on the x86-64 architecture.}
  \label{tab:design_fcd}
  \begin{tabular}{lc}
    \toprule
    & \textbf{Max. Call Depth} \\
    \midrule
    \textbf{500.perlbench\_r} & 310 \\
    \textbf{502.gcc\_r}       & \textbf{2581} \\
    \textbf{505.mcf\_r}       & 29 \\
    \textbf{520.omnetpp\_r}   & 303 \\
    \textbf{523.xalancbmk\_r} & 85 \\
    \textbf{525.x264\_r}      & 15 \\
    \textbf{531.deepsjeng\_r} & 63 \\
    \textbf{541.leela\_r}     & 779 \\
    \textbf{557.xz\_r}        & 20 \\
    \bottomrule
  \end{tabular}
\end{table}

By reducing the size of our return stacks, we directly strengthen the resilience of our dual stack scheme against brute-force probing.
For our scheme, the effective entropy, with which return stacks are placed in the program's address space, is only reduced by 3 bits, instead of 11 bits for SafeStack and AG-Stack.

\subsection{Return Stack Pointer}\label{sec:design_rsp}

Another design choice of SafeStack and AG-Stack is maintaining an architectural conformity (\threemark) by replacing the process' original stacks with safe stacks and using the architecture's regular stack pointer register to reference these stacks.
As discussed in \autoref{sec:dualstacks_attacks_leaks}, this results in pointer leaks through neglected metadata.
To avoid such undesired references to our return stacks, we opt to use a different register to hold the return stack pointer.

In particular, we choose one of the architecture's general-purpose registers and dedicate it exclusively to the return stack pointer.
As a consequence, in our dual stack scheme, the process' original stack becomes the unsafe stack and the architecture's regular stack pointer register still holds the unsafe stack pointer.
This way, we guarantee that neglected metadata may only contain references to the unsafe stack and that the return stack pointer cannot be leaked into unsafe memory.
This design choice also creates different compatibility options for the use with legacy (unprotected) libraries, which we will discuss in \autoref{sec:design_compatibility}.

\subsection{Return Stack Region}\label{sec:design_region}

As a consequence of \emph{not} replacing the process' original stacks with return stacks, we gain flexibility (\fourmark) when placing return stacks within the process' address space.
Our dual stack scheme is not forced to place the main thread's return stack at the system's designated memory location (i.e., \texttt{stack\_base}; see \autoref{fig:memory_layout}).
Instead, all return stacks can be allocated freely within the virtual address space.
This lifts the restriction on the entropy of the main thread's return stack, adjusting all return stacks in our scheme to the same entropy.
To specifically harden our dual stack scheme, we place all return stacks within a dedicated \emph{return stack region} that follows two design goals: (1)~Maintaining no metadata about its internal structure at any time and (2)~presenting a strong protection against information disclosure that severely impedes brute-force probing and directly renders memory allocation oracles impossible.

The return stack region is allocated in the memory mapping space using \smmap and occupies a size of $2^{44}$ bytes ($2^{32}$ pages).
Within this region, return stacks are assigned randomized base addresses resulting in an effective entropy of 29 bits ($2^{32}/2^3$).
Due to this randomness, we are further able to break any spatial adjacency that is otherwise experienced when using plain \smmap to allocate stacks within the memory mapping space.
Our return stack region guarantees stacks to have no spatial adjacency to themselves or to any well-known libraries that are stored in the memory mapping space.
Thus, we prevent an attacker from artificially constructing contiguous memory regions occupied by return stacks and from searching well-known libraries instead of the return stacks themselves.
As a consequence, the randomization within our return stack region greatly reduces the ability of an attacker to search for return stacks.
The effects of thread spraying are also minimized, as the return stack region is large enough to theoretically hold $(2^{32}-1)/9$ return stacks with page-sized gaps (i.e., guard pages) between them.
Summarizing, our dual stack scheme heavily impedes brute-force probing, even when supported by thread spraying.

To render memory allocation oracles impossible, and in contrast to other schemes that rely on information hiding, we remove all access permissions from the return stack region upon allocation.
Then, we only set read/write access on memory pages as needed, i.e., as soon as they are occupied by return stacks.
This way, we keep return stacks within an always-allocated, but non-accessible memory region.
Memory allocation oracles are therefore able to find the return stack region, but are unable to search for return stacks within that region.
To disclose stacks within the return stack region, an attacker has to perform an exhaustive search in a brute-force manner, which is heavily impeded as explained before.
Additionally, this design requires no metadata to manage safe stacks, and thus prevents the explicit leakage of stack locations.

\subsection{Stack Creation and Destruction}\label{sec:design_creation}

To manage the creation of new return stacks within the metadata-free return stack region, we rely on information about the access permissions of memory pages.
As no interface is exposed by the Linux kernel to retrieve this information, we leverage the \swrite syscall for this purpose.
The syscall takes a memory address and file descriptor as input and attempts to write the bytes stored at the specified address into the file.
On success, the syscall returns the number of bytes written indicating that the memory address is readable.
On failure, an error code is returned indicating no read permissions.
Note that \swrite only \emph{reads} from the specified address, and therefore has no effect on the process' memory.
Hence, we can utilize \swrite to determine whether memory pages of our return stack region are currently free (non-readable) or occupied by a return stack (readable), which enables us to create and destroy return stacks without maintaining any metadata.

For the stack creation, the return stack region is randomly probed using the \swrite syscall until 10 non-readable, consecutive memory pages (8 pages for the return stack, enclosed by two guard pages) are discovered.
As the region can hold a maximum of $(2^{32}-1)/9$ return stacks, a collision is highly unlikely.
Therefore, the performance overhead for stack allocations and the corresponding thread creations is negligible.
The permissions of the 8 center pages are set to read/write access and the dedicated return stack pointer register is initialized to the base address of the new return stack.
For the destruction of return stacks, the return stack pointer is obtained from its dedicated register and the permissions of the memory pages occupying the return stack are reset to non-accessible.

\subsection{Library Compatibility Options}\label{sec:design_compatibility}

Our dual stack scheme is designed to protect executables and libraries alike.
Depending on the usage scenario, it might be necessary to have interoperability with legacy libraries.
Therefore, we provide three options for library integration:
\begin{description}
	\item[Secure.]
    The library is compiled and instrumented with our dual stack scheme and therefore fully protected.
    Note that this option does \emph{not} require any changes to the source code of the library.
	\item[Aware.]
    The library is aware of the presence of return stacks without requiring instrumentation by our scheme.
    Because we utilize a general-purpose register to hold return stack pointers, libraries can be compiled to omit the usage of that register altogether.
    This allows us to link unprotected libraries into protected programs while guaranteeing the return stack pointer is never spilled by the unprotected library.
    As a result, our scheme facilitates the creation of fully aware runtime environments (e.g., an entire Linux system) that are able to execute protected and unprotected programs simultaneously without compromising the security of protected programs in any way.
	\item[Compatible.]
    A pre-compiled library is still compatible with our scheme.
    However, such a library might spill the return stack pointer onto the unsafe stack.
\end{description}
Summarizing, our dual stack scheme achieves compatibility to all types of libraries.
While the first two options require re-compilation, none of the cases require source code changes.

\section{Implementation}\label{sec:implementation}

We present implementations of our dual stack scheme for the x86-64 and ARM64 architectures.
The implementations are designed for the \glstext{ELF} binary format in conjunction with GNU/Linux execution environments.
Our dual stack scheme is integrated into the LLVM compiler framework (version 6.0.0), with modifications to the compiler backend (machine code generation) and the runtime support library.
For the execution of secured programs, our scheme further requires modifications to the GNU C standard and GCC runtime libraries.

\subsection{Compiler Backend}

Data on the stack is organized in stack frames---contiguous, per-function data regions---containing the function's return address and other stack objects, such as the function's parameters, register spills, and local variables.
Stack frames are created and destroyed by the prologue and epilogue of a function, which are emitted by the compiler backend during machine code generation.
To add support for our dual stack scheme, we modify function prologues and epilogues to maintain our return stack and store return addresses there instead of on the regular stack.
In the following, we present these modifications for the x86-64 and ARM64 \glspl{ISA}.

\subsubsection{x86-64 Architecture}

The x86-64 \gls{ISA} \cite{Intel-SDM} implements full descending stacks with the dedicated registers \texttt{RSP} to hold the stack pointer and \texttt{RBP} to hold the frame (or base) pointer.
While the stack is maintained by the function prologue and epilogue, the return address is pushed onto the stack by the \texttt{CALL} instruction and popped by the \texttt{RET} instruction.
In the following, we assume a typical function that saves one general-purpose register (\texttt{RBX}) alongside the frame pointer and return address onto the stack, and additionally reserves 64 bytes of stack space for local variables.
In x86-64 machine code, this translates to the prologue (ln. 1--4) and epilogue (ln. 6--9) shown on the left-hand side of \autoref{fig:asm_x64}.
The stack is extended by an additional 8 bytes, as \texttt{RSP} must be kept at an 16-byte alignment.
The return address is pushed onto the stack by the preceding \texttt{CALL} instruction, which is not depicted in the figure.

\begin{figure}
\begin{minipage}[t]{.44\linewidth}
\lstset{
  xleftmargin=15pt,
  xrightmargin=4pt
}
\begin{lstlisting}
 PUSH   %RBP
 MOV    %RSP, %RBP
 PUSH   %RBX
 SUB    $72, %RSP
 ...
 ADD    $72, %RSP
 POP    %RBX
 POP    %RBP
 RETQ
\end{lstlisting}
\end{minipage}
\hspace{.02\linewidth}
\begin{minipage}[t]{.52\linewidth}
 \lstset{
   xleftmargin=15pt,
   xrightmargin=4pt
}
\begin{lstlisting}
 (*\aftergroup\bfseries*)POPQ   (%R15)(*\aftergroup\mdseries*)
 (*\aftergroup\bfseries*)LEA    8(%R15), %R15(*\aftergroup\mdseries*)
 PUSH   %RBP
 MOV    %RSP, %RBP
 PUSH   %RBX
 SUB    (*\aftergroup\bfseries*)$64(*\aftergroup\mdseries*), %RSP
 ...
 ADD    (*\aftergroup\bfseries*)$64(*\aftergroup\mdseries*), %RSP
 POP    %RBX
 POP    %RBP
 (*\aftergroup\bfseries*)LEA    -8(%R15), %R15(*\aftergroup\mdseries*)
 (*\aftergroup\bfseries*)JMPQ   (%R15)(*\aftergroup\mdseries*)
\end{lstlisting}
\end{minipage}
\caption{Typical prologues and epilogues of a regular function (left) and a function with support for our return stack (right) on the x86-64 architecture.}
\label{fig:asm_x64}
\end{figure}

To extend the x86-64 \gls{ISA} for our dual stack scheme, an additional dedicated register has to be identified that holds the return stack pointer.
As discussed in \autoref{sec:design}, general-purpose registers are most suitable for this purpose, as they maintain compatibility with legacy libraries.
Based on this criterion, we choose \texttt{R15} as the dedicated register for the return stack pointer.

To support return stacks, it would be ideal to replace \texttt{CALL} instructions and push return addresses directly onto the return stack before jumping to the called function.
However, this is not an acceptable option, as it would break compatibility to legacy libraries that still use the \texttt{CALL} instruction.
Therefore, we opt to move return addresses from the regular stack onto the return stack during function prologues.
This is achieved through two additional instructions (ln. 1--2), as depicted on the right-hand side of \autoref{fig:asm_x64}.
The first instruction directly pops the return address from the regular stack onto the return stack, without using a temporary register.
The second instruction increments the return stack pointer by the size of the pushed return address.
The epilogue is extended by one additional instruction (ln. 11) to decrement the return stack pointer before returning by directly jumping to the address stored on the return stack (ln. 12).
We use the \texttt{LEA} instruction for manipulating the return stack pointer to ensure that the operation has no side effects on the flags register.
The remaining instructions (ln 3--10) are consistent with the regular prologue/epilogue pair, with the exception that no stack realignment has to be carried out as our prologue spills an even number of registers.

In summary, our implementation realizes an empty ascending stack to hold return addresses in only three additional instructions for each function's prologue/epilogue pair.

\subsubsection{ARM64 Architecture}

The ARM64 \gls{ISA}~\cite{ARM-v8} implements full descending stacks with the dedicated registers \texttt{SP} to hold the stack pointer and \texttt{FP} to hold the frame pointer.
In contrast to x86-64, the ARM64 \gls{ISA} does not provide a \texttt{CALL} instruction to push return addresses, but instead utilizes the so-called link register (\texttt{LR}) to pass them to called functions.
As a consequence, functions have to save return addresses themselves during the creation of stack frames.

In the following, we again assume a typical function that saves one general-purpose register (\texttt{X19}) alongside the frame pointer and return address onto the stack, and additionally reserves 64 bytes of stack space for local variables.
In ARM64 machine code, this translates to the prologue (ln. 1--4) and epilogue (ln. 6--9) shown in \autoref{fig:asm_aarch64_regular}.
The prologue first extends the stack by the combined size of register spills, alignment bytes, and local variables (i.e., $24+8+64$).
Then, the registers are spilled and the frame pointer is updated to point to the saved frame pointer value on the stack.
Likewise, the epilogue destroys the stack frame by reversing the prologue's
instructions.

\begin{figure}
\begin{lstlisting}
 SUB    SP, SP, #96
 STR    X19, [SP, #64]
 STP    FP, LR, [SP, #80]
 ADD    FP, SP, #80
 ...
 LDP    FP, LR, [SP, #80]
 LDR    X19, [SP, #64]
 ADD    SP, SP, #96
 RET
\end{lstlisting}
\caption{Typical prologue and epilogue of a function on the ARM64 architecture.}
\label{fig:asm_aarch64_regular}
\end{figure}

To extend the ARM64 \gls{ISA} for our dual stack scheme, we chose the general-purpose register \texttt{X28} to hold the return stack pointer.
Further, we implemented the empty ascending stack to hold return addresses by utilizing memory transfer instructions with pre- and post-indexed addressing unique to ARM architectures.
In pre-indexed addressing, the memory address accessed is the sum of the used base register and an offset, and the address is written back to the base register.
With post-indexed addressing, the accessed address is the value in the base register, and the sum of the address and the offset is written back to the base register.
Like this, we are able to add support for return stacks with \emph{zero or only two}\footnote{If the function spills an odd number of registers, we need zero additional instructions, if it spills an even number, we need two additional instructions, a store and a load.} additional instructions per prologue/epilogue pair.
Adapting the example prologue/epilogue pair from \autoref{fig:asm_aarch64_regular} requires no additional instructions.
The modifications are depicted in \autoref{fig:asm_aarch64_rs}.

The prologue first stores \texttt{LR} onto the return stack using a store instruction with post-indexed addressing (ln. 1).
The post-indexed store increments the base register \texttt{X28} by 8 bytes upon completion.
The remaining instructions (ln. 2--4), as in the regular prologue, reduce the stack pointer, spill the registers, and update the frame pointer.
Because our scheme stores the return address on the return stack, the prologue now spills an even amount of registers, namely \texttt{X19} and \texttt{FP}, which can be merged into one store instruction.
Additionally, no realignment is needed, which is why the stack is extended only by the size of register spills and local variables (i.e., $16+64$).
The epilogue (ln. 6--9) operates similarly to the regular epilogue, except that \texttt{LR} is restored from the return stack using a pre-indexed load instruction, which decrements \texttt{X28} by 8 bytes before loading the return address into \texttt{LR}.

\begin{figure}
\begin{lstlisting}
 (*\aftergroup\bfseries*)STR    LR, [X28], #8(*\aftergroup\mdseries*)
 SUB    SP, SP, (*\aftergroup\bfseries*)#80(*\aftergroup\mdseries*)
 (*\aftergroup\bfseries*)STP(*\aftergroup\mdseries*)    X19, (*\aftergroup\bfseries*)FP(*\aftergroup\mdseries*), [SP, #64]
 ADD    FP, SP, (*\aftergroup\bfseries*)#72(*\aftergroup\mdseries*)
 ...
 (*\aftergroup\bfseries*)LDP(*\aftergroup\mdseries*)    X19, (*\aftergroup\bfseries*)FP(*\aftergroup\mdseries*), [SP, #64]
 (*\aftergroup\bfseries*)LDR    LR, [X28], #-8!(*\aftergroup\mdseries*)
 ADD    SP, SP, (*\aftergroup\bfseries*)#80(*\aftergroup\mdseries*)
 RET
\end{lstlisting}
\caption{Typical prologue and epilogue of a function with support for our return stack on the ARM64 architecture.}
\label{fig:asm_aarch64_rs}
\end{figure}

\subsection{Runtime Support Library}

Our runtime support library adds a constructor to the initialization section of protected \glstext{ELF} binaries.
At load-time, the dynamic linker executes our constructor, which first initializes the return stack region using \smmap and then creates the return stack for the process' main thread.
The runtime support library also handles the creation and destruction of return stacks for multi-threaded \glstext{ELF} binaries.
For this, the thread initialization is intercepted.
Our interceptor function performs two tasks: Creating the child thread's return stack and registering a destructor to clean up the return stack upon thread cancellation.

\subsection{GNU/Linux Runtime Libraries}\label{sec:implementation_linux}

Executing protected \glstext{ELF} binaries under our dual stack scheme requires certain adjustments to the GNU/Linux runtime environment.
We discuss these adjustments in the following.

\subsubsection{Dynamic Linker}

For compatibility with our dual stack scheme, the dynamic linker of the GNU C standard library must be compiled with awareness of the dedicated return stack pointer register.
Because our runtime support library has to initialize the return stack pointer early in the loading process, the dynamic linker might clear our dedicated register at a later stage.
Therefore, we recompile the dynamic linker using the GCC compiler flags \texttt{-ffixed-r15} for x86-64 and \texttt{-ffixed-x28} for ARM64.
This guarantees the integrity of our return stack pointer register during load-time.

\subsubsection{Shortened Return Sequences}

As discussed in \autoref{sec:dualstacks_attacks_leaks}, the GNU C standard library spills registers to unsafe memory through the \setjmp/\longjmp feature.
To prevent leaks of our return stack pointer, we introduce functionally equivalent safe versions of \setjmp and \longjmp and add a custom transformation pass to the LLVM compiler framework.

The transformation pass performs two tasks: First, all calls to \setjmp and \longjmp are substituted for calls to their safe counterparts.
Second, for each call to \safesetjmp, a unique, pointer-sized marker is pushed onto the return stack.
To distinguish the marker from return addresses, a value above the process' maximum virtual address is used.
Next, \safesetjmp reads the marker upon invocation and stores it as part of the register state (i.e., instead of the return stack pointer).
Finally, when \safelongjmp is invoked, the marker is read from the register state and used to unwind the return stack.
The return stack is traversed backwards until the marker is encountered, decrementing the return stack pointer along the way.
This implicitly restores the original return stack pointer.

\subsubsection{Stack Unwinding}\label{sec:implementation_linux_unwinding}

As discussed in \autoref{sec:dualstacks_attacks_leaks}, the GCC runtime library needs to keep track of currently unwound stack frames.
This also applies to our return stack, as the unwinder eventually has to reinstate the correct return stack pointer.
To support our dual stack scheme without compromising its security, we extended the GCC runtime library to perform unwinding without storing the return stack pointer in the unwinding context.
Instead, we store an offset in the unwinding context, which indicates the position of the return address of the currently unwound stack frame relative to the top of the return stack.

On both, x86-64 and ARM64, stack unwinding instructions are generated according to the DWARF standard.
To implement return stack support with minimal ramifications on the existing unwinding specification, we introduce a new unwinding instruction, called \texttt{DW\_CFA\_def\_rsp\_offset}.
This unwinding instruction takes a single unsigned LEB128 offset representing the spill size of the stack frame (i.e., function) on the return stack (i.e., 8 bytes for the return address).
The instruction is inserted alongside the regular unwinding instructions in order to inform the unwinder to additionally unwind the return stack.
On encountering the instruction, the unwinder stores the offset in the unwinding context, adding it to any previously stored offset.
Like this, the unwinder accumulates the offset from the top of the return stack to the return address belonging to the currently unwound stack frame.
At the end, the unwinder directly updates the return stack pointer in its register by subtracting the offset.

\subsubsection{Multi-threading}

On GNU/Linux environments, support for multi-threading is provided by the GNU C standard library in form of user-level and kernel-level context switching.
User-level threads operate without kernel support and store switched contexts within the program's address space.
Therefore, user-level threads inherently leak the return stack pointer and must not be used when protecting programs with our dual stack scheme or any other equivalent scheme relying on information hiding.

On the other hand, kernel-level---or POSIX---threads rely on the Linux kernel to handle and store context switches.
As such, the return stack pointer is safely stored in the kernel's address space and our dual stack scheme is able to support POSIX threads naturally without leaking the return stack pointer.
Protected programs must be compiled with a modified version of the GCC runtime library (see \autoref{sec:implementation_linux_unwinding}), as the stacks of threads are forcefully unwound during thread cancellation to perform clean-up.

\section{Performance Evaluation}\label{sec:evaluation}

\newcommand{\meanintelss}{1.08\%\xspace}
\newcommand{\meanintelrs}{2.72\%\xspace}
\newcommand{\lossintel}{1.64\%\xspace}
\newcommand{\meanarmss}{0.50\%\xspace}
\newcommand{\meanarmrs}{-0.03\%\xspace}

We compare the performance of our leak-resilient dual stack scheme to a baseline with regular stacks and the SafeStack implementation in order to evaluate the overhead added by our additional security measures.
We primarily tested with the SPEC CPU2017 benchmark suite (with 2 iterations) on x86-64 and the EEMBC CoreMark benchmark (with 25,000 iterations) on ARM64.
We also applied both schemes to real world programs, i.e., the Apache (version 2.4.33) and Nginx (version 1.13.10) web servers, and tested their throughput with ApacheBench.
For this, we sent 1 million HTTP requests to our server from a remote (\glstext{LAN}) host and served each request with 128 bytes of data.
All evaluations were carried out on an Intel Core i5-7440HQ machine running Debian Buster and a Raspberry Pi 3 running Yocto Poky Linux.
The quad core processors of both systems were clocked at a static 2.2 GHz and 600 Mhz, respectively, to avoid frequency scaling.
For Debian Buster, we used dual stack-aware GNU C standard and GCC runtime libraries, while for the Raspberry Pi 3 we compiled a fully aware Yocto Poky Linux system from scratch (see \autoref{sec:design_compatibility}).
For all evaluations, we used natively LLVM-compiled binaries as baseline and enabled the compiler flags \texttt{-fsanitize=safe-stack} and \texttt{-fsanitize=return-stack} for the respective dual stack schemes.

\begin{table}
  \caption{Overhead of SafeStack and our leak-resilient dual stack scheme on the x86-64 \glstext{ISA}.}
  \label{tab:benchmark_intel}
  \begin{tabular}{l>{\centering\arraybackslash}m{1.25cm}>{\centering\arraybackslash}m{1.25cm}>{\centering\arraybackslash}m{2.1cm}}
    \toprule
    & \textbf{Baseline} & \textbf{SafeStack} & \textbf{Leak-Resilient Dual Stack} \\
    \midrule
    \textbf{500.perlbench\_r} &    570 s &  2.63\% &  4.21\% \\
    \textbf{502.gcc\_r}       &    411 s &  0.97\% &  5.60\% \\
    \textbf{505.mcf\_r}       &    623 s &  1.28\% &  1.44\% \\
    \textbf{520.omnetpp\_r}   &    687 s &  2.91\% &  2.04\% \\
    \textbf{523.xalancbmk\_r} &    570 s & -0.53\% &  0.18\% \\
    \textbf{525.x264\_r}      &    393 s &  0.51\% &  3.82\% \\
    \textbf{531.deepsjeng\_r} &    437 s &  1.14\% &  4.35\% \\
    \textbf{541.leela\_r}     &    761 s &  1.97\% &  7.10\% \\
    \textbf{557.xz\_r}        &    562 s &  0.71\% &  0.89\% \\
    \midrule
    \textbf{Apache}           & 130.70 s &  0.05\% &  0.11\% \\
    \textbf{Nginx}            & 178.54 s &  0.18\% &  0.15\% \\
    \midrule
    \textbf{Mean} & & \textbf{\meanintelss} &  \textbf{\meanintelrs} \\
    \bottomrule
  \end{tabular}
\end{table}

On the x86-64 \gls{ISA}, the performance measurements, as recorded in \autoref{tab:benchmark_intel}, show an average overhead of \meanintelrs for our scheme, compared to \meanintelss for SafeStack.
This means the additional security gained by our scheme comes at a mean loss of \lossintel in performance.
This is mainly due to the fact that our scheme adds three additional instructions to every function, whereas SafeStack only requires additional instructions for accessing objects on the unsafe stack, which does not occur for all functions.
However, even with our scheme, some programs, such as the Apache and Nginx web servers, as well as the \texttt{xalancbmk} and \texttt{xz} benchmarks, experience virtually no performance overhead, which we contribute to larger functions with fewer transitions in-between them.
This assumption is supported by our measurements of the maximum function call depth from \autoref{sec:design_size} (see \autoref{tab:design_fcd}), as programs with a shallow call depth will call fewer functions in general.

For the ARM64 \gls{ISA}, as recorded in \autoref{tab:benchmark_arm}, on average our scheme causes no measurable performance overhead (i.e., \meanarmrs), compared to an overhead of \meanarmss for SafeStack.
In comparison to x86-64, this is a clear performance gain for our scheme, which we attribute to the flexibility of the \gls{ISA}'s memory transfer instructions and the subsequent reduction in instructions needed per function.

In general, some benchmarks gain performance using our scheme, a result also observable in some benchmarks with SafeStack.
We attribute those results to improved caching behavior since both approaches can increase spatial locality of similarly accessed data, e.g., return addresses.

\begin{table}
  \caption{Overhead of SafeStack and our leak-resilient dual stack scheme on the ARM64 \glstext{ISA}.}
  \label{tab:benchmark_arm}
  \begin{tabular}{l>{\centering\arraybackslash}m{1.25cm}>{\centering\arraybackslash}m{1.25cm}>{\centering\arraybackslash}m{2.1cm}}
    \toprule
    & \textbf{Baseline} & \textbf{SafeStack} & \textbf{Leak-Resilient Dual Stack} \\
    \midrule
    \textbf{CoreMark} &  16.43 s & -0.73\% & -0.43\% \\
    \midrule
    \textbf{Apache}   & 196.36 s &  1.49\% &  0.85\% \\
    \textbf{Nginx}    & 215.98 s &  0.73\% & -0.50\% \\
    \midrule
    \textbf{Mean} & & \textbf{\meanarmss} & \textbf{\meanarmrs} \\
    \bottomrule
  \end{tabular}
\end{table}

\section{Related Work}\label{sec:rw}

To this day, a variety of defense mechanisms has been proposed to secure return addresses against exploitation through runtime attacks.
In the following, we briefly present defenses related to dual stack schemes and discuss their differences to our work.

Stack canaries \cite{CPM+98} are randomized values that are placed between the local variables and the return address of a stack frame and checked for their correctness before function returns.
Other designs refrain from adding canary values, but check the integrity of the saved frame pointer \cite{BST00,PHH+07} instead.
Stack canaries do not hold up against our threat model, as they only prohibit contiguous writes, but leave the stack unprotected against arbitrary writes.

Shadow stack schemes provide backward-edge \gls{CFI} by maintaining copies of return addresses in a separate memory region---the shadow stack---and using these copies to verify the integrity of function returns.
The different types of shadow stack schemes ensure the integrity of return addresses by either checking that the addresses on the regular stack match the stored copies \cite{CH01,PC03,GPSA06,NSW06,DSW11,KOAP12,ZQHS14,GCJ17}, or by overwriting the return addresses on the regular stack with the copies before issuing function returns \cite{ABEL05,DDE+12,DMW15}.
Shadow stacks only partly satisfy our threat model, as they detect both contiguous and arbitrary writes onto return addresses, but do not provide countermeasures against information disclosure attacks \cite{CCD+15}.
If anything, shadow stacks are typically stored at well-known memory locations (such as the data segment, within the \gls{TLS}, or at a fixed offset from the regular stack), enabling an attacker to overwrite the copies of return addresses.
Merely some recent hardware-assisted schemes deploy isolated monitoring processes, which inherently protects their shadow stacks from information disclosure, but also induces non-negligible performance overheads~\cite{GZZL17,LSW+17}.

Other defense solutions provide backward-edge \gls{CFI} based on heuristic schemes that either detect gadget chains based on their characteristics \cite{CZY+14,ZWS+14} or restrict function returns to call-preceded instructions \cite{XLCZ12,PPK13,ZS13,ZWC+13}, active call sites \cite{DHP+15}, or white-listed code locations \cite{GCJ17}.
However, heuristic-based defenses do not hold up against our threat model, as deviations from the intended \gls{CFG} are permitted within the scope of their respective heuristic \cite{GABP14,GAP+14,CBP+15}.

Dual stack schemes have been proposed even before SafeStack and AG-Stack.
SplitStack \cite{XKPI02}, ASR \cite{BSD05}, and XFI \cite{EAV+06} (with possible hardware support \cite{BEA06}) separate stack objects onto two independent stacks by applying different methods to differentiate between safe and unsafe objects.
Similar approaches apply source code transformations \cite{DM03,SGK05} to move unsafe objects to the heap instead of storing them on the stack.
Due to their similarities with SafeStack, these dual stack solutions suffer from the same weaknesses against information disclosure attacks.

True hardware-based shadow stack schemes provide superior resilience under our threat model, as the shadow stacks are completely out of the attackers reach by cryptographically securing them through hardware measures \cite{CLR05} or placing them in separate hardware caches \cite{LKMS04,OVB+06}.
However, these solutions require custom hardware rendering their application in off-the-shelf systems infeasible.
Outside of the academic research community, the need for backward-edge \gls{CFI} recently lead Intel and ARM to integrate practical solutions into their next generation processors.
Intel \glstext{CET} \cite{Intel-CET} introduces new \texttt{CALL} and \texttt{RET} instructions to maintain a shadow stack through a dedicated register and provides a new page table protection policy to secure the shadow stack in user-space memory.
The ARMv8.3 64-bit \gls{ISA} introduces new instructions that add cryptographic authentication codes \cite{Qualcomm-PAC} to pointers (including return addresses) stored in memory.
Both solutions provide superior resilience under our threat model but require the purchase of new hardware, not available in the near future.

Summarizing, in contrast to our presented dual stack design, none of the related approaches is able to provide a readily and widely usable, leak-resilient and secure solution for backward-edge \gls{CFI} that holds up against a strong threat model.

\section{Conclusion}\label{sec:conclusion}

In this paper, we presented a leak-resilient dual stack scheme capable of protecting return addresses even in the presence of sophisticated information disclosure attacks, including leveraging leaked pointers in libraries, probing in a brute-force manner, and utilizing memory allocation oracles.
To achieve this goal, we studied the vulnerabilities of previous dual stack designs and developed a novel approach for stack separation.
Our approach minimizes the size of the safe stack, carefully avoids spills of pointers to the safe stack and optimizes the random placement of the stack in memory.
Our approach is highly flexible and can be used to create \emph{completely} protected runtime environments that secure programs and libraries alike, but also to create \emph{aware} environments that are able to execute protected and unprotected programs simultaneously without compromising the security of protected programs in any way.
Our implementation using the LLVM compiler framework shows that our design is practical and highly efficient, on average causing no measurable performance overhead on ARM64 (i.e., 0.0\%) and only a negligible overhead of 2.7\% on x86-64.

\bibliographystyle{ACM-Reference-Format}

\end{document}